\documentclass[sigconf, screen]{acmart}

\usepackage{comment}
\usepackage{diagbox}
\usepackage{algorithm,algpseudocode, algorithmicx}
\algrenewcommand\alglinenumber[1]{\tiny #1:} 
\usepackage{setspace, etoolbox, caption}
\usepackage{graphicx}
\usepackage{epstopdf}
\usepackage{listings}
\usepackage{makecell,multirow}
\usepackage{url}
\usepackage{xcolor}
\usepackage{verbatim}
\usepackage{balance}
\usepackage{pifont}
\usepackage{enumitem}
\usepackage{adjustbox}
\usepackage{amsthm}
\usepackage{amsmath}
\usepackage{amsfonts}
\usepackage{textcomp,booktabs}
\usepackage[normalem]{ulem}
\usepackage{caption,subcaption}
\usepackage{tcolorbox}
\usepackage[T1]{fontenc}
\usepackage{threeparttable}
\usepackage{multicol}
\usepackage{flushend}

\newcommand{\tabincell}[2]{\begin{tabular}{@{}#1@{}}#2\end{tabular}}

\newcommand{\techname}{\textsc{HirGen}}
\newcommand{\bugsnum}{21}
\newcommand{\confirmedbugsnum}{17}
\newcommand{\confirmedandfixedbugsnum}{12}
\newcommand{\confirmedandunknownbugsnum}{10}
\newcommand{\fixedandunknownbugsnum}{5}
\newcommand{\unknownbugsnum}{14}
\newcommand{\fixedbugsnum}{12}
\newcommand{\opnum}{58}
\newcommand{\bopnum}{23}
\newcommand{\uopnum}{35}
\newcommand{\optimizationnum}{25}
\newcommand{\linenum}{3K}

\newcommand{\baselineNum}{four}

\newcommand{\eg}{\emph{e.g.}}
\newcommand{\ie}{i.e.}

\definecolor{sqc}{RGB}{255,187,0}

\usepackage{scalerel}
\usepackage{tikz}
\usetikzlibrary{svg.path}

\definecolor{orcidlogocol}{HTML}{A6CE39}
\tikzset{
  orcidlogo/.pic={
    \fill[orcidlogocol] svg{M256,128c0,70.7-57.3,128-128,128C57.3,256,0,198.7,0,128C0,57.3,57.3,0,128,0C198.7,0,256,57.3,256,128z};
    \fill[white] svg{M86.3,186.2H70.9V79.1h15.4v48.4V186.2z}
                 svg{M108.9,79.1h41.6c39.6,0,57,28.3,57,53.6c0,27.5-21.5,53.6-56.8,53.6h-41.8V79.1z M124.3,172.4h24.5c34.9,0,42.9-26.5,42.9-39.7c0-21.5-13.7-39.7-43.7-39.7h-23.7V172.4z}
                 svg{M88.7,56.8c0,5.5-4.5,10.1-10.1,10.1c-5.6,0-10.1-4.6-10.1-10.1c0-5.6,4.5-10.1,10.1-10.1C84.2,46.7,88.7,51.3,88.7,56.8z};
  }
}

\newcommand\orcidicon[1]{\href{https://orcid.org/#1}{\mbox{\scalerel*{
\begin{tikzpicture}[yscale=-1,transform shape]
\pic{orcidlogo};
\end{tikzpicture}
}{|}}}}

\algnewcommand\algorithmicswitch{\textbf{switch}}
\algnewcommand\algorithmiccase{\textbf{case}}
\algnewcommand\algorithmicassert{\texttt{assert}}
\algnewcommand\Assert[1]{\State \algorithmicassert(#1)}%
\algdef{SE}[SWITCH]{Switch}{EndSwitch}[1]{\algorithmicswitch\ #1\ \algorithmicdo}{\algorithmicend\ \algorithmicswitch}%
\algdef{SE}[CASE]{Case}{EndCase}[1]{\algorithmiccase\ #1}{\algorithmicend\ \algorithmiccase}%
\algtext*{EndSwitch}%
\algtext*{EndCase}%

\newcommand\myparagraph[1]{
 \vspace*{3pt}
  \noindent \textit{\textbf{#1.}}\quad
}

\begin{document}

\setcopyright{acmlicensed}
\acmPrice{15.00}
\acmDOI{10.1145/3597926.3598053}
\acmYear{2023}
\copyrightyear{2023}
\acmSubmissionID{issta23main-p76-p}
\acmISBN{979-8-4007-0221-1/23/07}
\acmConference[ISSTA '23]{Proceedings of the 32nd ACM SIGSOFT International Symposium on Software Testing and Analysis}{July 17--21, 2023}{Seattle, WA, USA}
\acmBooktitle{Proceedings of the 32nd ACM SIGSOFT International Symposium on Software Testing and Analysis (ISSTA '23), July 17--21, 2023, Seattle, WA, USA}
\received{2023-02-16}
\received[accepted]{2023-05-03}

\title{Fuzzing Deep Learning Compilers with \techname}

\author{Haoyang Ma}
\orcid{0000-0002-7411-9288}
\affiliation{
    \institution{Department of Computer Science and Engineering, The Hong Kong University of Science and Technology}
    \country{China}
}
\email{haoyang.ma@connect.ust.hk}

\author{Qingchao Shen}
\orcid{0000-0002-6128-2123}
\affiliation{
    \institution{College of Intelligence and Computing, Tianjin University}
    \country{China}
}
\email{qingchao@tju.edu.cn}

\author{Yongqiang Tian}
\orcid{0000-0003-1644-2965}
\affiliation{%
	\institution{University of Waterloo}
	\country{Canada}}
\affiliation{
\institution{The Hong Kong University of Science and Technology}
\country{China}}
\email{yongqiang.tian@uwaterloo.ca}

\author{Junjie Chen}
\orcid{0000-0003-3056-9962}
\affiliation{
    \institution{College of Intelligence and Computing, Tianjin University}
    \country{China}
}
\email{junjiechen@tju.edu.cn}

\author{Shing-Chi Cheung}
\orcid{0000-0002-3508-7172}
\authornote{Shing-Chi Cheung is the corresponding author.}
\affiliation{
    \institution{Department of Computer Science and Engineering, The Hong Kong University of Science and Technology}
    \country{China}
}
\email{scc@cse.ust.hk}

\begin{abstract}

Deep Learning (DL) compilers are widely adopted to optimize advanced DL models for efficient deployment on diverse hardware. Their quality has a profound effect on the quality of compiled DL models. A recent bug study shows that
the optimization of high-level \textit{intermediate representations} (IRs) is the most error-prone compilation stage
and bugs in this stage account for 44.92\% of the whole collected ones. 
However, existing testing techniques do not consider 
the features related to high-level optimization (e.g., the high-level IR), and are therefore weak in exposing bugs at this stage. 
To bridge this gap, we propose \techname{}, 
an automated testing technique that effectively exposes coding mistakes in the optimization of high-level IRs. 
The design of \techname{} includes 1) three coverage criteria to generate diverse and valid computational graphs; 2) the use of the high-level IR's language features to generate diverse IRs; 
3) three test oracles of which two are inspired by metamorphic testing and differential testing.
\techname{} has successfully detected \bugsnum{} bugs that occur at TVM, with \confirmedbugsnum{} bugs confirmed and \fixedbugsnum{} fixed. 
Further, we construct \baselineNum{} baselines using state-of-the-art DL compiler fuzzers that can cover the high-level optimization stage.
Our experiment results show that \techname{} can detect 10 crashes and inconsistencies that cannot be detected by the baselines in 48 hours. 
We also evaluate the usefulness of our proposed coverage criteria and test oracles.
\end{abstract} 



\maketitle

\section{introduction}
Deep learning (DL) compilers,
such as TVM~\cite{tvm}, Glow~\cite{glow}, XLA~\cite{xla} and nGraph~\cite{ngraph},
have shown effectiveness in optimizing advanced DL models 
for efficient model deployment at diverse devices~\cite{dlcsurvey}.  
They take as input a DL model, extract its computational graph, and re-represent the DL model
using intermediate representations (IRs)~\cite{dlcsurvey}.
DL compilers consist of multiple compilation stages, which include high-level and low-level optimizations. DL compilers arrange these two optimizations in order, with high-level optimization followed by low-level optimization.
The optimizations aim to compile deep learning models into binary executables 
that can run efficiently on target hardware devices.

Like conventional compilers\cite{sun2016toward,zhou2021empirical, jvmlanguagebugstudy, pyinterpreterstudy}, DL compilers are prone to bugs. 
These bugs can cause undesired compiler behaviors, such as crashes, 
unexpected wrong behavior,
and poor performance~\cite{qingchao}. These undesired behaviors
could result in catastrophic effects on the correctness and reliability of mission-critical DL applications (\eg, autonomous driving cars~\cite{selfdriving} and aircraft collision avoidance systems~\cite{aircraft}). 

Techniques have recently been proposed to detect bugs in DL compilers, 
including TZER~\cite{liu2022coverage}, TVMFuzz~\cite{TVMFuzz}, MT-DLComp~\cite{MTXiao} and NNSmith~\cite{NNSmith}.
Despite preliminary reported success in bug detection for TVM, 
they are inefficient in revealing bugs that occur in high-level optimization, which account for 44.92\% of the bugs found in DL compilers~\cite{qingchao}.
TZER and TVMFuzz~\cite{liu2022coverage, TVMFuzz} are proposed to detect low-level optimization bugs in a DL compiler with generated low-level IRs. 
Since these two techniques test DL compilers by mutating low-level IRs, and low-level IRs cannot be used by high-level optimization, they theoretically cannot detect bugs in the high-level optimization stage. 
MT-DLComp~\cite{MTXiao} tests a DL compiler by constructing mutated DL models. 
Since its mutation strategies only insert operators that yield zero, the kinds of operators and the available places to insert these operators are limited. Therefore, it cannot generate models of diverse computational graphs to cover corner high-level optimization cases.
In test oracle design, MT-DLComp does not take advantage of the language features of the high-level IR and high-level optimizations.
As a result, it cannot effectively detect bugs in high-level optimizations as demonstrated in Section \ref{sec:5}.
NNSmith~\cite{NNSmith} mainly focuses on revealing the hidden defects in DL compilers, such as arithmetic problems, fragile type systems, and poor support for specific data layouts, by generating computational graphs and inputs. 
Since its generation process and design of test oracle do not consider language features of the high-level IR in DL model generation, and do not consider high-level optimizations in test oracle design, it cannot efficiently detect bugs
related to high-level optimization. 
In Section~\ref{sec:5}, we will show NNSmith is orthogonal to \techname{} regarding bug detection ability.


\myparagraph{\techname{}}
To bridge the gap, 
we propose the first DL compiler fuzzing technique that focuses on high-level optimization, \techname{}. 
\techname{} is designed to satisfy the following four objectives: 
1) the satisfaction of integrity constraints, such as type match and tensor shape match, that govern high-level IRs to avoid an early crash before invoking optimization, 
2) the exploration of the diversity of computational graphs, 
3) the utilization of the high-level IR's language features for the construction of diverse high-level IRs,
4) the capability of detecting multiple types of optimization bugs.
To achieve the first objective, 
\techname{} performs type checking and shape checking in each insertion of the operator node by leveraging the information of each existing node, including its type, shape, and connections. After insertion, \techname{} also updates the information of the new node for future use. 
To meet the second objective, \techname{} is coverage-guided in input space to explore diverse operator nodes, operator edges, and the combination of operator type and data type.
To meet the third objective, \techname{} can construct diverse high-level IRs from a single computational graph to achieve full use of the IR's language features.
To meet the fourth objective, \techname{} incorporates three test oracles, two of which are designed purposely for DL compilers.
An example of test oracles is that a model should not make a different prediction after optimization. 
Besides functional correctness, \techname{} can also test the robustness of DL compilers.
Specifically, \techname{} provides an option of generating invalid computational graphs that violate type constraints and shape constraints~\cite{docter}. 
This option tests whether DL compilers can catch such invalid computational graphs and throw the expected exceptions. This way, \techname{} can also detect bugs caused by incorrect exception handling.


Following prior works on DL compiler testing, we evaluate the performance of \techname{} on TVM, which is the most popular DL compiler. 
For baseline selection, we choose 1) TVMfuzz (with lowercase f), a preliminary proof-of-concept application from a bug study~\cite{qingchao}. The tool is chosen because it is the only testing technique that focuses on detecting bugs arising from high-level optimizations in DL compilers; 
2) MT-DLComp~\cite{MTXiao}, a metamorphic testing framework that can cover the high-level optimization stage; 
3) LEMON~\cite{LEMON}, a fuzzing technique for DL libraries (e.g., Tensorflow~\cite{tensorflow}, PyTorch~\cite{pytorch}) testing; 
and 4) NNSmith~\cite{NNSmith}, a generation-based DL compiler fuzzer.
We repeated the comparison experiment ten times to mitigate the influence of randomness.
Our experimental results show that 1) \techname{} can detect around ten distinct crashes that are not detectable by other techniques in a two-day execution; 2) TVMfuzz, MT-DLComp, and LEMON are all inefficient in detecting bugs, they found about three crashes in total; 3) NNSmith is also efficient in bug detection, detecting nine distinct crashes. But except for one crash, all the crashes that they found are orthogonal to the crashes found by \techname{}. 
We will elaborate on the experimental results in Section \ref{SOTA}.
In addition to this comparison experiment, we examine the usefulness of the coverage-guided strategy in generating diverse computational graphs by an ablation study in Section \ref{sec:RQ5}. 

\myparagraph{Contribution} 
This study makes three major contributions. 
\begin{itemize}[leftmargin=8pt, topsep=0pt]
    \item This work introduces a new focus on testing the most bug-prone stage, high-level optimization, of DL compilers. We propose a computational graph generation algorithm and three test oracles to detect bugs of diverse root causes in high-level optimizations. 
    \item We implement \techname{}, a fuzzing technique targeting at TVM. 
    It has detected \bugsnum{} bugs, of which
    \confirmedbugsnum{} have been confirmed, 
    \fixedbugsnum{} have been fixed, and 
    \unknownbugsnum{} bugs were previously unknown. 
    Among the \confirmedbugsnum{} confirmed bugs, 14 are highly related to high-level optimizations, and others are about low-level optimization and deployable code generation.
    Furthermore, our extensive evaluation shows that \techname{} 
    outperform 
    the state-of-the-art testing techniques.
    \item We release \techname{}, the details of detected bugs and experiment data at \url{https://github.com/haoyang9804/HirGen}.
\end{itemize}

\section{Background}

\subsection{DL Compilers}
DL compiler takes as input DL models. These models can be constructed with the help of DL frameworks, such as Tensorflow\cite{tensorflow} and PyTorch\cite{pytorch}. 
After interpreting the computational graph of input model, 
DL compiler converts it into a high-level IR. 
Each node in the computational graph is represented by one or several IR expressions. For instance, a \textit{conv2d} (two-dimensional convolution) node is represented by $nn.conv2d$ in the high-level IR of TVM.
DL compilers then optimize the computational graph at the high-level IR. 
For instance, a static subgraph independent of inputs can be optimized through constant folding at the IR. 
After optimization, the high-level IR is translated into the low-level IR for further optimization. 
In this step, a high-level IR expression (e.g., {\tt nn.conv2d}) 
is expanded into a nested loop of low-level computation instructions. 
Subsequently, low-level optimizations are performed to improve efficiency. 
For instance, loop tiling can be performed at low-level optimization to accelerate the computation of $conv2d$ on a specified hardware device. 
Finally, the low-level IR is translated into deployable code for diverse hardware using traditional compilers and platforms. 
\subsection{Computational Graph and High-Level IR}

\begin{figure}[ht]
    \centering
    \begin{adjustbox}{center}
    \setlength{\tabcolsep}{0.2em} 
    {\renewcommand{\arraystretch}{1.0}
    \begin{tabular}{cc}
    \begin{subfigure}{.25\textwidth}
        \vspace{-3mm}
        \begin{adjustbox}{center}
        \includegraphics[scale=0.5]{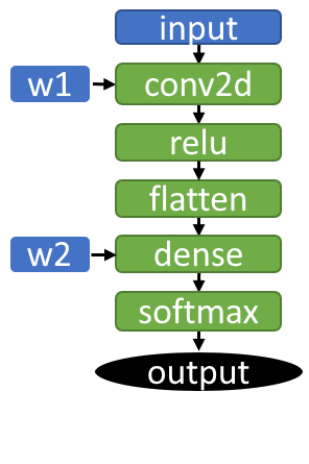}
        \end{adjustbox}
        \vspace{-3mm}
        \caption{Computational Graph}
        \vspace{-2mm}
        \label{figure:cg}
    \end{subfigure}
    &
    \begin{subfigure}{.25\textwidth}
        \vspace{-3mm}
        \begin{adjustbox}{center}
        \includegraphics[scale=0.5]{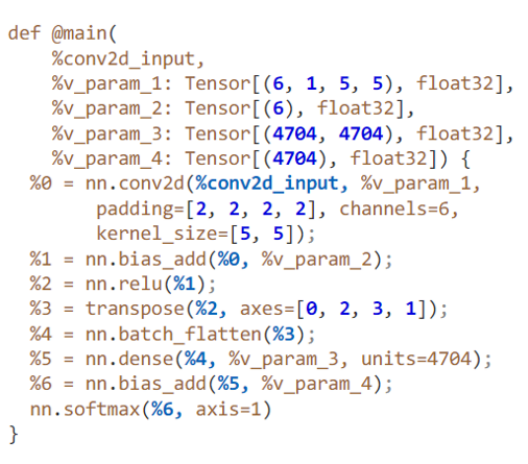}
        \end{adjustbox}
        \vspace{-3mm}
        \caption{High-level IR}
        \vspace{-2mm}
        \label{figure:hir}
    \end{subfigure}
    \end{tabular}}
    \end{adjustbox}
    \caption{An Example of a Computational Graph and the Corresponding High-level IR}
    \vspace{-1mm}
    \label{figure:cgir}
\end{figure}

A computational graph is a directed graph that expresses the data flow in computation. 
High-level IR, also known as graph-level IR, 
is an intermediate representation to express the computational graph. 
In DL community, high-level IRs are widely used to describe computational graphs by DL compilers (e.g., TVM) and also frameworks (e.g., ONNX). 
Figure \ref{figure:cg} illustrates the computational graph of a 2-dimensional convolutional neural network, where variable/constant nodes and operator nodes are colored blue and green, respectively. 
A black ellipse denotes the end of a graph. 
Arrows in this graph represent data flows. 
Specifically, variable and constant nodes are the starting points of a data flow, passing their data to the following nodes. 
In contrast, operator nodes work as a relay to extend the data flow, passing the results they calculate. 
Figure \ref{figure:hir} shows the corresponding
Relay IR, 
the high-level IR in TVM, of the computational graph in Figure \ref{figure:cg}. This IR is not unique as the same semantics can be represented in other ways 
by utilizing the language features of IR. 
For instance, Relay offers first-class functions to separate the main function into multiple functions with complex call chains. 
ONNX also plans to support this feature in the future~\cite{onnxoperators}.

\section{Approach}
\label{sec:3}
\begin{figure}
\centering
    \includegraphics[width=0.90\linewidth]{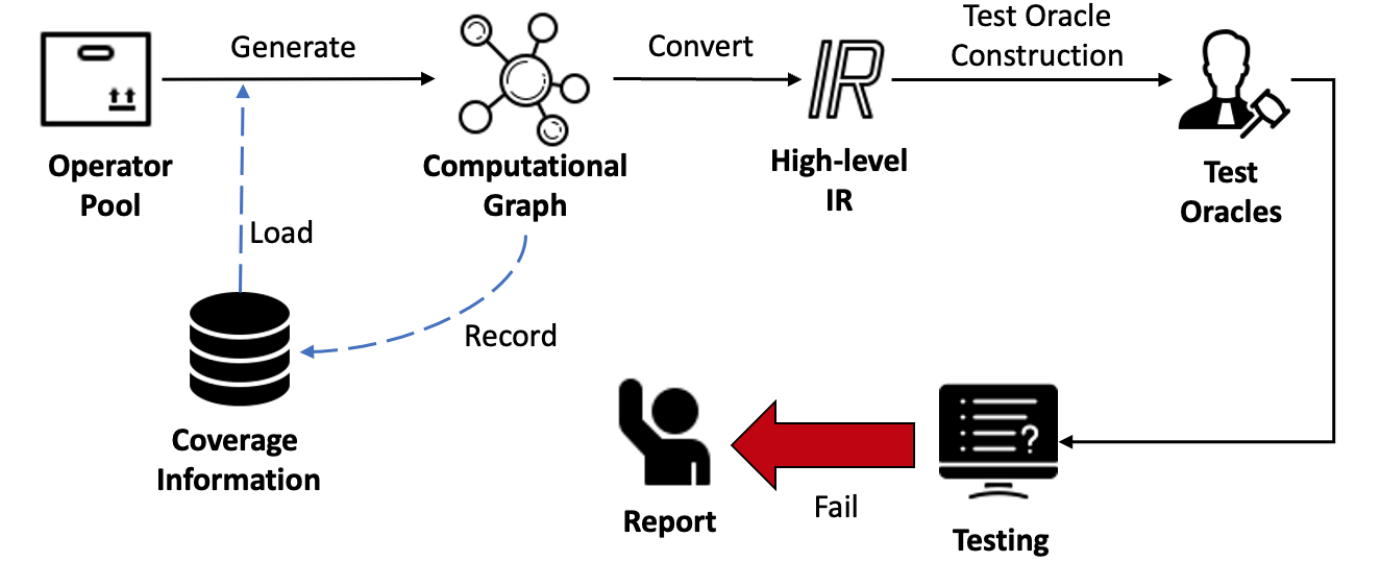}
    \caption{Workflow of \techname{}
    } 
    \label{figure:overview}
\end{figure}

This section presents the design and underlying methodology of \techname{}.
Figure \ref{figure:overview}
shows the workflow of \techname{}. 
\techname{} maintains a pool of \opnum{} operators that can be expressed by high-level IRs in popular high-level frameworks (e.g. Relay~\cite{relay}, ONNX~\cite{onnx}).
\techname{} first loads existing coverage information, generates a computational graph based on it and updates coverage information from the newest graph. Then, \techname{} leverages a high-level framework such as Relay or ONNX, 
to convert the graph into a high-level IR and feeds it into the DL compiler. 
To efficiently capture the defects in the target DL compiler, 
besides the commonly-used oracle crash, 
\techname{} constructs two test oracles from the spirits of metamorphic testing and differential testing.
Any test case that violates the oracles is regarded as a witness to a bug of the compiler and will be reported to developers. 
The remainder of this section is divided into three parts. 
Section \ref{subsection:3-1} elaborates on the details of our computational graph generation algorithm. 
Section \ref{subsection:3-2} presents how to utilize the language features of the high-level IR and convert computational graphs into high-level IRs.
In Section \ref{subsection:3-3}, we will introduce the design of the test oracles proposed by us.

\subsection{Computational Graph Generation}
\label{subsection:3-1}

\subsubsection{Overview.}

We consider the generation of a computational graph as a process of continuously inserting various operator nodes into the initially empty graph $CG=\left\{\right\}$ until the number of operator nodes 
reaches a threshold.
In each iteration, \techname{} selects an operator from the operator pool, loads the operator into $CG$ as node $nd$, and constructs a connection between $nd$ and other existing nodes. 
Meanwhile, \techname{} maintains the information
of each node, including data type, tensor shape, and connection information. 
Furthermore, \techname{} also involves three coverage criteria
to improve the diversity of the graph.

With these prerequisites, \techname{} provides two generation modes. To generate valid computational graphs, \techname{} utilizes the aforementioned node information for strict type checking and shape checking. 
In this way, each insertion strictly satisfies the constraints. 
We call this mode \textit{strict generation}.
However,
strictly following the constraints may miss the opportunity to test the exception handling ability of DL compilers when constraints are violated.
Therefore, \techname{} also provides \textit{disruptive generation} to deliberately break type constraints and shape constraints.
We will first elaborate on the strict generation and then the disruptive generation.


\subsubsection{Node Information}
Inserting a node to $CG$ requires node information of all existing nodes to perform type-checking and shape-checking.
%
Node information describes the typical features of a node, and such information is essential to ensure the correctness of each insertion.
For instance, in the insertion of an operator named {\tt add} that sums two nodes, \techname{} first checks all available nodes (including operator nodes, variable nodes, and constant nodes) and selects two nodes $n_a$ and $n_b$ from available nodes,
such that $n_a$ and $n_b$ have compatible tensor shapes and data types, and their data types are acceptable by the operator {\tt add}.
Each type of node has its own node information, as detailed in Table~\ref{tab:info}.

\begin{table}[ht]
\centering
\caption{Node Information}
\resizebox{\linewidth}{!}{
\begin{tabular}{c|l}

\textbf{Node Type} & \textbf{Node Information}\\
\midrule 
$variable$ & $dataType$, $tensorShape$ \\
$constant$ & $dataType$, $tensorShape$, $value$ \\
$operator$ & $dataType$, $parentNodes$, $tensorShape = $ \textproc{inference}($parentNodes$) \\
\bottomrule
\end{tabular}
}

\label{tab:info}
\end{table}

Specifically, \techname{} considers the following three types of nodes.
\begin{enumerate}[leftmargin=12pt,topsep=0pt]
    \item \textit{Variable} node. It involves data type $dataType$ and tensor shape $tensorShape$ describing the details of the tensor wrapped in this node. 
    $dataType$ corresponds to the data type of all elements in this tensor, such as $int64$ and $float32$.
    $tensorShape$ is a vector of the scale of all dimensions in the tensor.

    \item \textit{Constant} node. 
    Besides the $dataType$ and $tensorShape$, \textit{constant} node includes the value of tensor $value$ as a part of its information. 
    \item \textit{Operator} node. 
    Operators require parameter(s); thus, they are all connected with other nodes in the graph.
    To document this connection information for each operator node, 
    besides $dataType$, \techname{} records its parent node(s) $parentNodes$ to which this node connects and records its tensor shape inferred from the parent node(s). 
\end{enumerate}

\subsubsection{Coverage Guidance.}
\label{sec:coverageguidance}
To explore diverse data types, shapes, and operators in computational graph generation, we design three coverage criteria.
    
\myparagraph{Operator-datatype Coverage}
Let $op_i$ be the $i_{th}$ operator in the operator pool. Let $dtype_j$ be the $j_{th}$ data type in the collection of data types.
$Cov(op_i, dtype_j)$ is $1$ when $op_i$ has once been inserted into the graph as a node with data type $dtype_j$. Otherwise, it is $0$.
This coverage encourages \techname{} to 1) involve different operators in the graph and 2) utilize diverse data types since the data type problem is a big concern for DL compilers~\cite{qingchao}.

\myparagraph{Operator-shape Coverage}
Let $op_i$ be the $i_{th}$ operator in the operator pool. Let $shape$ be the shape of the output tensor of this operator node after being inserted into the graph. Let $Cov(op_i, shape)$ be $1$ if $op_i$ has once been inserted into the graph as a node with tensor shape $shape$, and 0 otherwise.
With operator-shape coverage, \techname{} tries various calculations with diverse tensor shapes, thus increasing the probability of encountering calculation problems, such as the poor implementation of some operators in special shapes or different calculation results on different platforms.

\myparagraph{Operator-edge Coverage}
Let $op_i$ and $op_j$ be the $i_{th}$ and $j_{th}$ operators in the operator pool. Let $Cov(op_i, op_j)$ be $1$ if there exists one edge from $op_i$ to $op_j$, and $0$ otherwise.
This coverage guides \techname{} to connect the new operator node to the existing operator nodes instead of variable and constant nodes. This way, the generated computational graph contains a more complex and deep data flow instead of a parallel connection of several simple data flows.
In the implementation, we can easily extend operator-edge coverage to operator-path coverage with three or more operator nodes included. In this way, we can explore a more diverse and complicated computational graph, but at greater time costs.

The design of the first two coverage criteria is motivated by the fact that type problem and shape problem are the two major root causes of DL compiler bugs~\cite{qingchao}. 
The design of the third one tries to complicate the data flow of the computational graph since the third coverage encourages \techname{} to interleave different operators in a computational graph.
Specifically,
With these three coverage criteria, \techname{} is prevented from producing previously generated subgraphs in computational graph generation. Since high-level optimizations often involve identifying, annotating, re-constructing, and shrinking optimizable subgraphs, duplicate subgraphs can only find duplicate bugs.
For instance, \textit{operator fusion} is a high-level optimization that can fuse operators in a high-level expression into a larger operator. 
Fusing different operators could encounter different situations and cover different codes.
With coverage criteria, a group of operators will unlikely appear in a high-level expression in the previous order. 
Therefore, these three coverage criteria facilitate \techname{} to find new bugs more efficiently. $Bug_2$ in Table \ref{tab:bugs} was detected by \techname{} in three days but was never detected by \techname{}$_r$ (\techname{} without coverage guidance) in our 2-week experiment.

\subsubsection{Constraint-Awareness Graph Generation}

Algorithm \ref{Algorithm1} presents two procedures used by \techname{} to generate computational graphs that strictly follow the type and shape constraints of the high-level IR. 
\textproc{generation} is the main procedure. 
It takes as input the required number of operators $rOpNum$ to be contained in the output computational graph.
\textproc{preinsert} is an auxiliary procedure that enforces 
type-checking and shape-checking in generation.
\textproc{generation} procedure includes the following two main parts. 
\begin{algorithm}[t]
\footnotesize
\caption{Computational Graph Generation}
\label{Algorithm1}
\begin{algorithmic}[1]
\Procedure {Generation}{$rOpNum$}
    \State $CG \leftarrow \{\}$ \label{a1:initbeg}
    \State $opnum \leftarrow 0$ 
    \State $opPool \leftarrow \{add, subtract, multiply, divide, ...\}$
    \State $dataTypeSet \leftarrow \{int64,int32,int16,int8,uint64,uint32,...\}$\label{a1:initend}
    \Repeat \label{a1:itbeg}
        \State $opNode \leftarrow$ \textproc{select}($opPool$) \label{a1:selectop} 
        \State $dataType \leftarrow$ \textproc{select}($dataTypeSet$) \label{a1:selectdtype}
        \State $connection, shape, CG \leftarrow$ \textproc{preInsert}($opNode$, $dataType$, $CG$) \label{a1: connection_shape}
        \State $coverage \leftarrow$ \textproc{GetCoverage}($opNode$, $dataType$, $connection$, $shape$) \label{a1: cov}
        \If{\textproc{newCoverage}(coverage)}
            \State $opnum \leftarrow opnum + 1$
            \State \textproc{UpdateCoverage}(coverage)
            \State $opNode.info \leftarrow (connection, shape, dataType)$
            \State $CG \leftarrow CG \bigcup \{node\}$
        \EndIf \label{a1: finisupdate}
    \Until{$opnum$ = $rOpNum$} \label{a1:itend}
    \State \Return $CG$
\EndProcedure

\Procedure {preInsert}{$opNode$, $dataType$, $CG$}
    \State $availableNodes \leftarrow$ \textproc{typeCheck}($CG$, $dataType$) \label{a1: typecheck}
    \State $nodeGroup1, nodeGroup2, ... \leftarrow$ \textproc{shapeCheck}($availableNodes$) \label{a1: shapecheck}
    \State $paramNodes \leftarrow$ \textproc{select} ($nodeGroup1, nodeGroup2, ... $) \label{a1: dump}
    \If{\textproc{nodeNotEnough}($paramNodes$)} \label{a1: begnodenotenough}
        \State $node1, node2, ... \leftarrow$ \textproc{create}($dataType$, $paramNodes$)
        \State $CG \leftarrow CG \bigcup \{node1, node2, ...\}$
        \State $paramNodes \leftarrow paramNodes \bigcup \{node1, node2, ...\}$
    \EndIf \label{a1: endnodenotenough}
    \For {$node$ \textproc{in} $paramNodes$} \label{a1:begconnection}
        \State $connection \leftarrow (opNode, node)$
    \EndFor \label{a1:endconection}
    \State $shape \leftarrow$ \textproc{Inference}($connection$)
    \State \Return $connection, shape, CG$
\EndProcedure

\end{algorithmic}
\end{algorithm}

\myparagraph{Initialization} 
\techname{} performs initialization from Line~\ref{a1:initbeg} to Line ~\ref{a1:initend}. Specifically, an empty computational graph $CG$ is initialized and the number of operators $opnum$ in $CG$ is set to $0$. The operator pool and data type set are both initialized for future use.

\myparagraph{Generation Loop} 
In each iteration (Lines \ref{a1:itbeg}-\ref{a1:itend}), \techname{} generates an operator node, updates its node information, and finally inserts it into $CG$ if new coverage is explored. 
Specifically, \techname{} first randomly selects an operator and data type (Line \ref{a1:selectop} and \ref{a1:selectdtype}). Then it seeks for connection from $CG$ and infers the tensor shape of the operator node $opNode$ built from the newly selected operator (Line \ref{a1: connection_shape}). 
Subsequently, \techname{} calculates coverage and performs update and insertion if new coverage is explored (Line \ref{a1: cov}-\ref{a1: finisupdate})
\ie, there is any increment of the three coverages defined in section \ref{sec:coverageguidance}.
During the update, \techname{} adds $opnum$ by $1$, updates coverage and node information of $opNode$.
When $opnum$ equals $rOpNum$, \techname{}
stops the generation loop and
returns $CG$.

Procedure \textproc{preInsert} shows the details of building connection(s) between $opNode$ and existing nodes of $CG$ and shape inference. 
In type-checking, \techname{} absorbs type constraints in the target DL compiler and uses them to avoid type mismatch (e.g., float32 and int64 for \textit{add} operator). 
In shape-checking, \techname{} includes shape rules in the target DL compiler, such as broadcasting rule, which specifies that if the two arrays differ in their number of dimensions, the shape of the one with fewer dimensions is padded with the ones on its leading side. Shape-checking avoids shape mismatch.
With type-checking (Line \ref{a1: typecheck}) and shape-checking (Line \ref{a1: shapecheck}), \techname{} sorts out several node groups from $CG$. All nodes of each node group are mutually shape-compatible and nodes from these groups are all type-compatible with $opNode$. Then \techname{} selects a node group and dumps all its nodes into $paramNodes$ (Line \ref{a1: dump}). 
The number of required parameter nodes is fixed for each kind of operator. In the implementation, \techname{} has a certain probability of connecting one node to an operator node multiple times, such as connecting one variable node to add a node twice, meaning adding the node to itself. But to complicate the data flow, \techname{} discourages this behavior in favor of connecting the required number of different nodes.
Therefore, if the number of parameter nodes in $paramNodes$ is insufficient for $opNode$,
\techname{} has a large probability of creating variable nodes or constant nodes that are shape-compatible with all parameter nodes and type-compatible with $opNode$, inserts them into $CG$ and updates $paramNode$ (Line \ref{a1: begnodenotenough}-\ref{a1: endnodenotenough}). Finally, \techname{} creates connection information (Line \ref{a1:begconnection}-\ref{a1:endconection}), infers the tensor shape of $opNode$ and returns them both plus the possibly updated $CG$.

\subsubsection{Disruptive Generation Algorithm.} 
Disruptive generation is similar to constraint-awareness generation. It also needs coverage to memorize what type constraints and shape constraints have been broken. In addition, node information is also required. Since it contains the data type and tensor shape of each node, which is necessary for breaking constraints.
Specifically, during disruptive generation, \techname{} purposely 1) connects the operator node to other node(s) with the data type(s) it cannot accept in TVM (e.g., \texttt{add} operator with {\tt bool} data type), and 2) connects nodes that are type-incompatible or shape-incompatible (e.g., add two nodes of which the shapes are {\tt [3,4]} and {\tt [2,3]} respectively). 

\subsection{High-Level IR Generation}
\label{subsection:3-2}

High-level IR generation is simple with the help of existing high-level frameworks, such as Relay and ONNX.
Taking Relay as an example, it provides ample APIs for receiving node information of various types of nodes and diverse operator nodes. For instance, {\tt relay.var} takes as inputs its name, data type, and tensor shape, and {\tt relay.add} takes as input only its connection information. These APIs contain strict type constraints and shape constraints, and it is easy to crash early before optimization if the computational graph contains an error. 

Besides the plain conversion by loading each node into its corresponding high-level expressions and assembling them into a high-level IR, we can also utilize the primitive features of these high-level frameworks. 
Take Relay for example, to improve expressivity, it allows using a function to wrap a subgraph and call the function in other ones. ONNX also plans to support this feature by supporting Function API. To better utilize these features, we also consider extracting a subgraph from the generated computational graph and wrapping it with a high-level function. In this way, we can better test how DL compilers tackle the situation where functions are included.

\begin{algorithm}[t]
\footnotesize
\caption{Conversion from a Computational Graph to a High-level IR}
\label{Algorithm2}
\begin{algorithmic}[1]
\Procedure {Conversion}{$CG$}
    \State $Functions \leftarrow \{\}$ \label{a2:init1}
    \State $Expressions \leftarrow \{\}$ \label{a2:init2}
    \For {$node$ \textproc{in} $CG$} \label{a2:begfor}
        \State $Expressions \leftarrow$ \textproc{load}($node.info$, $Functions$, $Expressions$) \label{a2:load}
        \If{\textproc{roll}() $ == func$} \label{a2:begif}
            \State $inputNodes, outputNodes \leftarrow $ \textproc{analyze}($Expressions$) \label{a2:inputoutput}
            \State $function \leftarrow$ \textproc{composeFunction} ($inputNodes, outputNodes$) \label{a2:composefunction}
            \State $Functions \leftarrow Functions \bigcup function$  \label{a2:updatefunc}
            \State $Expressions \leftarrow \{\}$ \label{a2:updateexpr}
        \EndIf \label{a2:endif}
    \EndFor \label{a2:endfor} 
    \State \Return $Expressions \bigcup Functions$
\EndProcedure
\Procedure{load}{$node.info$, $Functions$, $Expressions$}
    \State $expression \leftarrow$ \textproc{ConstructExpression} ($node.info$) \label{a2:ConstructExpression}
    \If {\textproc{parentInFunction}($node.info$)} \label{a2:parentinfunction}
        \State $functions \leftarrow$ \textproc{find}($node.info$) 
        \State $callExprs \leftarrow$ \textproc{createCallExpression}($functions$) \label{a2:callexprcreate}
        \State $Expressions \leftarrow Expressions \bigcup \{callExprs\}$
    \EndIf
    \State $Expressions \leftarrow Expressions \bigcup \{expression\}$
    \State \Return $Expressions$
\EndProcedure
\end{algorithmic}
\end{algorithm}


The overall algorithm of converting a computational graph into a high-level IR is shown in Algorithm \ref{Algorithm2}. 
\textproc{Conversion} procedure takes as input a computational graph $CG$ and outputs its corresponding high-level IR.
During initialization, \techname{} creates two empty sets, named $Functions$ and $Expressions$ respectively (Line \ref{a2:init1}, \ref{a2:init2}). They represent the collection of functions and high-level expressions, respectively. In the for loop (Line \ref{a2:begfor}-\ref{a2:endfor}), \techname{} traverses all nodes in $CG$, loads each node into high-level expression and update $Expressions$ (Line \ref{a2:load}). It randomly selects a set of high-level expressions and wraps them with a function (Line \ref{a2:begif}-\ref{a2:endif}). To compose a function, \techname{} first analyzes the input nodes and output nodes of the underlying subgraph of the expressions (Line \ref{a2:inputoutput}). It composes a function using these nodes (Line \ref{a2:composefunction}. Finally, \techname{} updates $Functions$ and $Expressions$ (Line \ref{a2:updatefunc}-\ref{a2:updateexpr}).
\textproc{Conversion} procedure returns the union of $Expressions$ and $Functions$ as the high-level IR.
\textproc{load} procedure presents the detail of loading a node into high-level expression. During loading, \techname{} takes care of connection information by inquiring whether the node connects to other nodes wrapped in function(s) (Line \ref{a2:parentinfunction}), if it is the case, then a call expression is created (Line \ref{a2:callexprcreate}). This procedure returns $Expressions$ after the update.

\subsection{Test Oracles}
\label{subsection:3-3}
Test oracles determine if a test passes or fails.
In this paper, we consider three test oracles to detect different types of failures.
Any failed test case determined by these oracles will be reported.

\subsubsection{\textbf{$Oracle_1$}: \textit{Crash.}}
Crash is widely used in test oracle construction to decide whether the testing fails \cite{liu2022coverage}. 
Besides, according to the statistics in a compiler bug study \cite{qingchao}, the number of bugs with the crash symptom occupy $59.37\%$ of all collected $603$ bugs. This huge proportion shows an urgent need to take crashes seriously. 
As for crash bugs detected when type-checking and shape-checking are turned off, we only report the bug if the crash is a segmentation fault because other crashes with detailed bug traces are primarily due to explicit violations of constraints in the computational graph. 
As for other crash bugs, we report them all since the generated computational graph under checking strictly follows all constraints in TVM and the crash is largely due to the poor implementation of TVM.

\subsubsection{\textbf{$Oracle_2$}: \textit{Result Inconsistency among the Original High-Level IR, the Optimized High-Level IR and the Mutated High-Level IR}}
Intuitively, high-level optimization only relates to performance boosts such as calculation acceleration and memory cost saving, but can not change results. 
In addition to involving high-level optimization, we also design a mutation strategy named function rewrite to generate the mutated high-level IRs that have the same output as the original high-level IR given the same input. 
This mutation strategy is inspired by Relay's support for functional programming features. By function rewriting, we can better utilize Relay's expressions and better test TVM with richer high-level IRs.
Specifically, this mutation strategy can rewrite function expressions in the high-level IR in the following ways.
\begin{itemize}[leftmargin=10pt, topsep=0pt]
    \item Turn a global function $f$ into the local closure of another newly created global function $g$. $g$ has the same parameters as $f$ and its returned value is a call to $f$ with these parameters. After tuning, this mutation also substitutes all calls to $f$ with calls to $g$.
    \item Wrap a function $f$ with an empty function $g$ which returns $f$ and also change all calls to $f$ to calls to the call to $g$.
    \item Call a function $f$ and return the call in another function $g$, then substitute all calls to $f$ with calls to $g$.
\end{itemize}

The mutated high-level IR only differs from the original high-level IR in the function call chain. 
Therefore, it is expected that the three high-level IRs produce the same calculation results given the same input. This metamorphic relation inspires us to form this oracle.
In addition to different calculation results, if the original high-level IR passes compilation and runtime but the optimized or mutated one fails in one of these two processes, we also count it as the result inconsistency.

\subsubsection{\textbf{$Oracle_3$}: \textit{Result Inconsistency across Hardware Devices}}
To maintain the same predictive capability of a DL model on different supported hardware devices, TVM should promise to output the same results on diverse hardware given the same input to a DL model. And similar to $Oracle_2$, inconsistent execution status (e.g., crash on CPU but execute well on GPU) is also counted as result inconsistency.
Following this common sense, we build $Oracle_3$ with the spirit of different testing. 
Given any high-level IR, after compiling it with multiple provided compilation approaches, feeding input and executing it on CPU and GPU, it is reasonable to expect the same calculation results.

\section{Experiment Setup}
\label{sec:4}

\subsection{Research Questions}

In this study, we aim to answer the following research questions:
\begin{description}[leftmargin=10pt]
    \item [RQ1] How effective is \techname{} in detecting bugs of TVM?
    \item [RQ2] Are all the test oracles effective in detecting bugs?
    \item [RQ3] Are bugs found by \techname{} highly related to high-level optimization?
    \item [RQ4] Is disruptive generation useful in finding exception-handling bugs?
    \item [RQ5] Can coverage-guided generation benefit the diversity of the computational graph?
\end{description}

\subsection{\techname{} Implementation}

We implement \techname{} in C++ with around \linenum{} lines of code.
Our implementation involves \opnum{} operators~\cite{HirGen-operator} to generate computational graphs, \optimizationnum{} high-level optimizations for catch optimization bugs and four compilation methods to conduct testing.

\vspace{-2mm}
\subsubsection{Operators}
In total, \techname{} includes \opnum{} operators supported by TVM, including \bopnum{} binary operators and \uopnum{} unary operators.
And it is easy to extend \techname{} with other operators.

\vspace{-2mm}
\subsubsection{Optimization and Compilation Methods.}

We select in total \optimizationnum{} high-level optimizations supported by TVM~\cite{HirGen-optimization}.
The main reason for choosing these high-level optimizations in TVM is our generated computational graphs can trigger them. 
Besides these high-level optimizations, it is easy to extend \techname{} with other optimizations.
Besides collecting these high-level optimizations, we also utilize different compilation methods provided by TVM. Different compilation methods deal with different scenarios and include different optimization sequences.
Overall, \techname{} supports the following four compilation methods.
\begin{enumerate}[topsep=0pt,leftmargin=15pt]
    \item {\tt relay.build()}
    \item {\tt relay.build\_module.create\_executor(`debug')}
    \item {\tt relay.build\_module.create\_executor(`graph')}
    \item {\tt relay.build\_module.create\_executor(`vm')}
\end{enumerate}

\subsection{Bug Report}
For each bug we have found, we report it in one of the three channels: 
1) upload the bug-triggered script and experiment environment on TVM Community~\cite{TVM-Community};
2) report the bug on Github Issue~\cite{TVM-Issue} with a reproducible script, experimental environment, and most importantly, our analysis on the reason for triggering it; 
3) create a pull request with the elaboration of this bug and our code patch. 
We choose our reporting channels primarily based on our expertise in the problem.
For the least familiar bug, we submit it on TVM Community in the form of a question to get rid of misdiagnosis. 
Then, we wait for an official fix or some comments from developers on this problem. 
For the most familiar one, we directly fix it, and we succeed in creating two pull requests and fixing two bugs. 
For other situations, we choose the second way and leave some comments on how to fix the bug.

\subsection{Baseline Selection}

We selected four baselines from the literature.

\myparagraph{TVMfuzz}
TVMfuzz is a preliminary proof-of-concept application for fuzzing TVM~\cite{qingchao}.
It can learn TVM API call chains from unit test scripts, then re-order and mutate them. By learning from high-level IRs and optimization-related unit test scripts, TVMfuzz can cover this stage.

\myparagraph{MT-DLComp}
MT-DLComp is an automated testing framework for DL compilers~\cite{MTXiao}. 
It mutates existing DL models to generate equivalent models and test DL compilers by three oracles. 
Though this technique is not specially created for detecting bugs in high-level optimization, it can cover this bug-prone stage. Therefore, we also include it as a baseline.

\myparagraph{LEMON}
LEMON is a testing technique for deep learning frameworks~\cite{LEMON}. 
It generates Keras~\cite{keras} models by mutating existing models. By setting different backends of Keras, LEMON detects prediction differences incurred by these backends.
Though LEMON is not for testing DL compilers, we can retrofit it to barely achieve the goal. In short, we remain the mutation part to generate new models and test DL compilers by two test oracles: 1) crash and 2) above-threshold prediction difference between original Keras models and compiled Keras models.

\myparagraph{NNSmith}
NNSmith is a generation-based fuzzer for DL compilers~\cite{NNSmith}.
During generation,
it generates diverse computational graphs, converts them into DL models using different DL frameworks, and uses gradient-guided search to generate inputs. 
During testing, 
it conducts differential testing among several DL compilers. 
In the testing process, NNSmith captures all prediction differences and crashes. 


\subsection{Metrics}
We mainly target \textit{bug counting} for evaluation.
To evaluate \techname{}, we count bugs based on independent fixes and developers' confirmation in Section \ref{sec:5.1.1}.
In Section \ref{SOTA} and \ref{sec:RQ5}, we studied five baselines in total and obtained a number of crashes/inconsistencies. 
Since many of them are duplicates, reporting them to the TVM community for bug confirmation can be time-consuming and possibly receives no reply according to our interaction with the developers. 
Therefore, we use the proximity of bug counting in the experiment of these two sections. In particular, manually-deduplicated bugs in Section 5.1.1 have totally different stack traces, and thus comparing the number of crashes/inconsistencies with distinct stack traces is reasonable. The proximity of bug counting is also used in the evaluation of other works~\cite{csmith, FuzzMNN}.


\subsection{Miscellaneous}
\myparagraph{Timeout Setting}
There are two comparison experiments involving timeout. The first is comparing \techname{} with the four baselines. The second is comparing \techname{} with \techname{}$_r$.
    We executed each of the involved techniques separately for two days, and each execution was conducted ten times to mitigate the influence of randomness. 
    Since all the studied techniques do not find distinct crashes/inconsistencies after 26 hours, it indicates that our 2-day timeout is reasonable to a large extent.

\myparagraph{Platform}
We conducted experiments on a server with Intel Xeon CPU, NVIDIA GeForce GTX1080Ti GPU, and 128 RAM, coordinated with 64-bit Ubuntu 16.04 OS.

\section{Evaluation}
\label{sec:5}

\subsection{RQ1: Bug Detection Capability of \techname{}}
\label{sec:5.1.1}
Running three months under the strict mode and one week under the disruptive mode, \techname{} has found \bugsnum{} bugs, of which \confirmedbugsnum{} have been confirmed.  
\confirmedandfixedbugsnum{} out of \confirmedbugsnum{} confirmed bugs have been fixed. 
Moreover, \confirmedandunknownbugsnum{} bugs are previously unknown and \fixedandunknownbugsnum{} fixed bugs were previously unknown.
Table \ref{tab:bugs} presents the details of all the confirmed bugs discovered by \techname{}, including their symptoms, root causes, the test oracles detecting them, the fixing status, whether they are previously unknown, by which generation mode were they detected, were they also found by other techniques (blanks mean no other techniques detected the bugs in experiments) and whether they are high-level optimization bugs.
Symptom includes crash and inconsistency. 
The former means that TVM terminates unexpectedly while the latter means that different results or statuses are caught in testing. 
We also manually investigate the root cause of each bug
adopting the taxonomy of a recent bug study~\cite{qingchao}.
Specifically, We carefully compare these bugs with the collected historical bugs and assign each of them a root cause. 

There are five root causes resulting in these bugs.
\begin{itemize}[leftmargin=8pt, topsep=0pt]
    \item Type Problem. This category of bugs is triggered by data type-related problems, including incorrect type inference, incomplete implementation of an operator on one data type, etc.
    \item Incorrect Exception Handling. This category of bugs occurs when TVM lacks rich and readable warning messages or even has no handling of some extreme situations. This kind of bug is related to the robustness of TVM.
    \item Incorrect Numerical Computation. This root cause involves incorrect numerical computations, values, or usages.
    \item Internal API Incompatibility. This category of bugs is triggered because TVM can not handle the combination of some APIs correctly. For instance, unexpected refusal of one combination of several high-level optimizations is counted as this kind of bug.
    \item Memory Allocation Problem. 
    This root cause refers to poor or incorrect memory allocation.
\end{itemize}

Check marks in \textit{Previously Unknown} column in Table~\ref{tab:bugs} indicate that the corresponding bug was unknown before we reported it. Since TVM was evolving fast, we found some cases early in the experiment that crashed on the version we tested (TVM v0.9, commit id: 124813f) but worked fine on the latest version. These bugs have been actually fixed before being reported and thus marked as previously known bugs.

\begin{table*}
\centering
\footnotesize
\caption{Confirmed Bugs found by \techname{}}
\resizebox{\linewidth}{!}{%
\begin{tabular}{@{}c|c|c|c|c|c|c|c|c@{}}
\toprule
\multirow{2}{*}{Bug ID} & \multirow{2}{*}{\textbf{Symptom}} & \multirow{2}{*}{\textbf{Root Cause}}          & \textbf{Test} & \multirow{2}{*}{\textbf{Status}} & \textbf{Previously} & \textbf{Generation} & \textbf{Found} & \textbf{High-level} \\
 &  &           & \textbf{Oracle} & & \textbf{Unknown} & \textbf{Mode} & \textbf{By} & \textbf{Optimization} \\
\midrule
1           & Crash            & Incorrect Numerical Computation               & $Oracle_1$              & Fixed           & \checkmark & strict & NNSmith &                                     \\
2           & Inconsistency    & Incorrect Exception Handling & $Oracle_2$              & Confirmed           & \checkmark   &  strict  & & \checkmark                              \\
3           & Crash            & Incorrect Exception Handling & $Oracle_1$              & Fixed           & \checkmark   &  strict   &  & \checkmark                                  \\
4           & Crash            & Incorrect Exception Handling & $Oracle_1$              & Fixed           &         &  strict    & &                            \\
5           & Crash    & Incorrect Exception Handling                 & $Oracle_1$              & Fixed           & \checkmark        &  strict     & &                             \\
6           & Inconsistency    & Incorrect Exception Handling & $Oracle_2$              & Fixed           & \checkmark              &  strict & & \checkmark                         \\
7           & Inconsistency    & Type Problem                 & $Oracle_2$              & Fixed           &                      &  strict    & & \checkmark              \\
8           & Inconsistency    & Type Problem                 & $Oracle_2$              & Fixed           &                         &  strict   & & \checkmark            \\
9           & Inconsistency    & Internal API Incompatibility & $Oracle_2$              & Fixed           & \checkmark                &  strict   & & \checkmark                     \\
10          & Inconsistency    & Incorrect Exception Handling & $Oracle_2$              & Fixed           &                            &  strict   & & \checkmark \         \\
11          & Crash            & Incorrect Exception Handling                 & $Oracle_1$              & Fixed           &    &  disruptive & & \checkmark \\
12          & Crash            & Incorrect Exception Handling                 & $Oracle_1$              & Fixed           &   &  disruptive & & \checkmark  \\
13          & Crash            & Incorrect Exception Handling                 & $Oracle_1$              & Fixed           &   &  disruptive & & \checkmark  \\
14          & Crash    & Memory Allocation Problem  & $Oracle_1$              & Confirmed           &  \checkmark            &  strict & & \checkmark \\
15          & Inconsistency    &  \tabincell{l}{Incorrect Numerical Computation}  & $Oracle_3$              & Confirmed           &  \checkmark  &  strict & &         \\
16          & Inconsistency    & \tabincell{l}{Incorrect Numerical Computation} & $Oracle_3$              & Confirmed           &  \checkmark   &  strict & &       \\
17          & Inconsistency    & \tabincell{l}{Incorrect Numerical Computation} & $Oracle_3$              & Confirmed           &  \checkmark   &  strict   & &     \\
\bottomrule
\end{tabular}
}
\label{tab:bugs}
\begin{tablenotes}
\footnotesize
\item Empty cells in column \textbf{Previously Unknown} refer to the bugs that are previously known; empty cells in column \textbf{Found by} refer to the bugs that are not found by other techniques; empty cells in column \textbf{High-level Optimization} refer to the bugs that are not relevant to high-level optimization.
\end{tablenotes}
\end{table*}

\myparagraph{Comparison with State-of-the-art Techniques}
\label{SOTA}
On average, \techname{} detected $11.8$ distinct crashes/inconsistencies. The variance of the number of them in the 10 repeated experiments is $0.36$. We also conducted a manual inspection of these crashes/inconsistencies by two experienced researchers. We observed that the average number of crashes/inconsistencies related to high-level optimization is $8.8$, 
and the variance of the number is $0.36$.
TVMfuzz detected $3.7$ distinct crashes on average, of which $1.4$ crashes are related to high-level optimizations. By Mann-Whitney U Test~\cite{evaluatefuzz}, $p$-value of the difference between \techname{} and TVMfuzz is $0.00018 < 0.01$, which implies the result that \techname{} outperforms TVMfuzz in DL compiler bug detection has statistical significance.
MT-DLComp and LEMON do not detect any bugs. 
As for NNSmith, it detected $10$ distinct crashes/inconsistencies on average.
Among these crashes/inconsistencies, data layout problems and data type problems are predominant, altogether accounting for $52.2\%$ of all crashes/inconsistencies. 
They are captured with bug messages such as \textit{"WCHN layout is not supported"}
or \textit{"TVM cannot support type matching between int32 and int64"}. 
Among other crashes/inconsistencies, on average $3.5$ crashes/inconsistencies are related to high-level optimization, and the variance is $1.45$, showing that NNSmith is unstable in detecting high-level optimization bugs. 
The $p$-value of the high-level optimization crashes detection difference between \techname{} and NNSmith is also $0.00018 < 0.01$, implying the result that \techname{} outperforms NNSmith in high-level optimization bug detection has statistical significance.
During the manual inspection, we only found one overlapping crash detected by both \techname{} and NNSmith, showing that these two techniques have almost complementary bug detection abilities. We will discuss the reason in Section \ref{sec:NNSmith}.

\subsection{RQ2: Effectiveness of Test Oracles}
\label{sec:RQ2}
To demonstrate the effectiveness of our test oracles, we conduct a case study of several representative and confirmed bugs detected by each test oracle.

\subsubsection{$Oracle_1$: \textit{Crash}} 
$Oracle_1$ caught the most bugs among all test oracles. 
In total, it finds eight bugs with three root causes, including \textit{Incorrect Numerical Computation}, \textit{Incorrect Exception Handling}, \textit{Memory Allocation Problem}.


\myparagraph{Incorrect Numerical Computation} 
Take $Bug_1$ as an example. 
In the computational graph that triggers this bug, a {\tt divide} operator first calculates the result of dividing a constant by a variable and then passes the calculation result $R$ to {\tt floor\_mod} as a dividend.
All involved variable nodes and constant nodes are of data type $uint$ and this type finally flows into {\tt floor\_mod}. 
However, TVM pre-calculates the possible value range of $R$ and detects it could probably be 0. 
Therefore, TVM incorrectly throws an exception and terminates even before we give values to $var1$ and $var2$, This bug only happens when the data type is $uint$ and is caused by incorrect value range estimation. 
After developers confirmed this bug and fixed {\tt const\_int\_bound} analyzer, this numerical computation-related bug was fixed.

\myparagraph{Incorrect Exception Handling} $Bug_{11}$, $Bug_{12}$ and $Bug_{13}$ are three bugs of Incorrect Exception Handling. They are detected under disruptive generation. 
To trigger these bugs, \techname{} must generate computational graphs containing obvious breaks of constraints. 
For example, in $Bug_{11}$, 
the bug-triggering computational graph includes a constant node of type $int16$, a {\tt tan} operator node and the connection between these two nodes. 
The constant node passes its $int16$ data to the operator node.
In this graph,
\techname{} purposely breaks the constraint that {\tt tan} only accepts $float$ data type defined in TVM and receives a segmentation fault during compilation. 
This is because TVM does not have exception handling for this operator and its unacceptable data types.

\myparagraph{Memory Allocation Problem}
$Bug_{14}$ is the only bug of this root cause. Specifically, when \techname{} leverages {\tt relay.shape\_of} to infer the tensor shape of the variable node with static tensor shape $(1, 2)$, an unexpected crash happens with warning message {\tt Cannot allocate memory symbolic tensor shape [?, ?]}. The question mark here refers to a dynamic shape.

\subsubsection{$Oracle_2$: \textit{Result Inconsistency among the Original High-Level IR, the Optimized High-Level IR, and the Mutated High-Level IR}}
$Oracle_2$ caught a total of six confirmed bugs, and five of them have been fixed. 
These bugs are caused by three different root causes, including \textit{Incorrect Exception Handling}, \textit{Type Problem}, and \textit{Internal API Incompatibility}.

\myparagraph{Incorrect Exception Handling}
Take $Bug_{10}$ as an example.
\techname{} catches this bug because it finds that a high-level IR passes compilation while its optimized version fails. 
Specifically, \techname{}
places {\tt FirstOrderGradient} before {\tt FuseOps} in an optimization sequence and detects that TVM cannot successfully handle this optimization sequence.
This is because exception handling is too strict. 
Concretely, TVM performs a traversal on the high-level IR after {\tt FirstOrderGradient} for conducting {\tt FuseOps}. 
When visiting a constant node, TVM finds this node is not scalar because {\tt FirstOrderGradient} has rewritten this attribute. Therefore, TVM throws an exception and the compilation terminates. 
However, this check about scalar attributes is too strict and does not consider data type. 
A fix for this bug completes this exception handling and makes the optimized version successfully passes compilation. 
Besides, $Bug_6$ is also a representative, detected by our effort in utilizing the high-level IR's language features. \techname{} takes advantage of first-citizen functions in Relay IR and tries to return a function in another function. Since TVM v0.9 cannot well support the lowering of this high-level language feature into a low-level counterpart, a segmentation fault is thrown. The effort in utilizing the high-level IR's language features also helps us find $Bug_5$, $Bug_7$, and $Bug_8$.

\myparagraph{Type Problem}
Take $Bug_8$ for instance. 
This bug is detected by function rewrite mutation. Specifically, after changing a global function $f$ into the local closure of another empty global function $g$ and returning $f$ in $g$, TVM can not infer the type of $g$. 
This is because after successfully inferring the type of $f$, this type information is lost when TVM begins to infer the type of $g$. 

\myparagraph{Internal API Incompatibility}
The bug $Bug_9$  is detected because
{\tt relay.build\_module.create\_executor(’vm’)}
fails, but compilation in other ways runs smoothly. 
Specifically, after \techname{} transforms a high-level IR into the A norm form.
Compilation with the virtual machine cannot figure out the bound relation between $x_{91}$ and a global function. 
However, other compilation ways do not encounter this problem. 

\subsubsection{$Oracle_3$: \textit{Result Inconsistency across Hardware Devices}}
$Oracle_3$ caught a total of three confirmed bugs, but none of them has been fixed. 
This is because the difference between computation results on CPU and GPU is caused by platform-specific differences. 
More specifically, LLVM and CUDA have different implementations of the same operator, while TVM lacks full specification about this operator or lacks a complete warning message about using this operator. 
Developers responded with a confirmation of this deficiency but they consider it unnecessary to remedy it without it violating the effectiveness of TVM seriously.

Take $Bug_{15}$ as an example. 
\techname{} creates a simple computational graph containing a {\tt right\_shift} operator node. This operator node takes as input two other variable nodes.
Subsequently, \techname{} first generates the corresponding high-level IR, then compiles the IR with {\tt relay.build} to generate the runtime model, and finally creates the input and runs the runtime model on CPU and GPU to get two computation results. 
When the second variable is larger than the first one, the results are inconsistent. 
This is because this situation incurs a poison value in LLVM and the use of it in an operator is undefined. 
Although this confirmed bug does not come from a bad implementation of TVM but from an external compiler issue, it still confuses users when their DL model triggers this inconsistency.
The refinement of the exception handling system could be a compromise approach for this ill situation.

\subsection{RQ3: Bugs Related to High-Level Optimization}
\label{RQ3}
As a DL compiler fuzzer focusing on high-level optimization, \techname{} is capable of detecting bugs in high-level optimization or bugs highly related to this stage. 
In this subsection, we manually study the code patch of each fixed bug detected by \techname{} and analyze their relationship with high-level optimization and how the detection of them improves this stage.

$Bug_2$, $Bug_8$, $Bug_9$, $Bug_{10}$ are bugs detected in high-level optimization. Bug-triggered patterns for these four bugs are similar: after high-level optimizations, \techname{} detects a violation of $Oracle_2$. These bugs show the inability to optimize the structure that several high-level optimizations should have optimized, and incompatibility among several optimizations. For instance, $Bug_8$ shows that after performing {\tt InferType} on one function, the solved types cannot be passed to the next function and thus triggers a type problem. $Bug_{10}$ shows {\tt FuseOp} can not be well performed after performing {\tt FirstOrderGradient}. 
Fixing these bugs directly 
improves the performance of the optimization and facilitates the possibility of multiple optimization combinations. 

Besides, \techname{} finds eight bugs with crash symptoms and all of them were triggered during compilation. Among them, $Bug_3$, $Bug_{14}$ are directly related to high-level optimization.
To improve efficiency, TVM calls {\tt OptimizeImpl} during compilation and invokes 11 high-level optimizations implicitly. 
These optimizations work by one or several passes on the high-level IR, which performs a rewrite at any optimizable expression. 
All expressions in the high-level IR are visited in each pass, and assertions embedded in TVM check each expression. 
Bugs in this process may prevent high-level optimizations from being well executed or even result in a crash to stop the optimization. Fixes for these bugs are indirect fixes for the required IR passes needed by high-level optimizations. 
Besides, $Bug_{11}$, $Bug_{12}$, and $Bug_{13}$ are in the high-level IR construction. Since construction happens before optimization, these bugs also prevent high-level optimizations.

Although our approach is proposed for high-level optimization, the test cases generated by our approach can also execute low-level optimizations and deployable code generation. Thus, it has the side effect of testing the other stages. The results also confirm it.
$Bug_{15}$, $Bug_{16}$, and $Bug_{17}$ are all related to the low-level part and code generation of TVM. They are detected due to inconsistent calculation results on different backends (i.e., LLVM and CUDA) given the same inputs. These bugs show the need to couple TVM with these backends better. $Bug_1$ and $Bug_5$ are arithmetic problems at low-level. \techname{} can detect them because the generated computational graphs contain error-triggered computational logic.

\subsection{RQ4: Effectiveness of Disruptive Generation}
\label{sec:RQ4}
During experiments, \techname{} generated 170 computational graphs with different bug-triggering combinations of the operator, data type, and tensor shape. All these graphs can incur crashes of TVM with only ``segmentation fault'' information, showing the deficiency of exception handling ability. In the latest TVM version, all these bugs have been fixed. All these obvious breaks of constraints trigger crashes with detailed bug information now. By comparing the bug information of the latest TVM, we found there are three bugs in total triggered by these 170 graphs.

\subsection{RQ5: Effectiveness of Coverage-Guided Generation}
\label{sec:RQ5}
To generate diverse computational graphs with various data types, tensor shapes, and operators, we design three coverage criteria. To answer this RQ,
We implement a simplified version of \techname{}, saying \techname{}$_{r}$. \techname{}$_{r}$ is identical to \techname{} except that \techname{}$_{r}$ is not guided to generate computational graphs but selects operators, shapes, and types under the rule that all selections are valid, and connects the new operator to the existing operator(s) randomly under the rule that the connection is valid. We conducted two experiments for the comparison between these two techniques.



The first experiment is about the bug detection ability of \techname{} and \techname{}$_{r}$ under strict generation mode. In each round of the experiment, we executed these two techniques for two days independently. To mitigate the influence of randomness, our experiment includes ten rounds.
On average, \techname{} found $11.8$ distinct crashes/inconsistencies. Among them, $8.8$ crashes/inconsistencies are related to high-level optimization. The variance of the number of distinct crashes/inconsistencies and that of the number of distinct crashes/inconsistencies related to high-level optimizations are both $0.36$.
As for 
\techname{}$_{r}$, it found $8.9$ distinct crashes/inconsistencies. $6.7$ crashes/inconsistencies are related to high-level optimizations. The variance of the number of distinct crashes/inconsistencies is $3.16$ and the variance of the number of distinct crashes/inconsistencies related to high-level optimizations is $1.12$. 
By Mann-Whitney U Test, the $p$-value of \techname{} outperforming \techname{}$_r$ in detecting distinct crashes/inconsistencies related to high-level optimizations is $0.0028 < 0.01$, and
the $p-value$ of \techname{} outperforming \techname{}$_r$ in detecting distinct crashes/inconsistencies is $0.00512 < 0.01$, implying the result that \techname{} outperforms \techname{}$_r$ in detecting high-level optimization bugs has statistical significance.
Besides, \techname{}$_r$ has a bigger variance, showing that it's unstable in bug detection.

\begin{figure}
     \centering
    \includegraphics[width=0.85\linewidth]{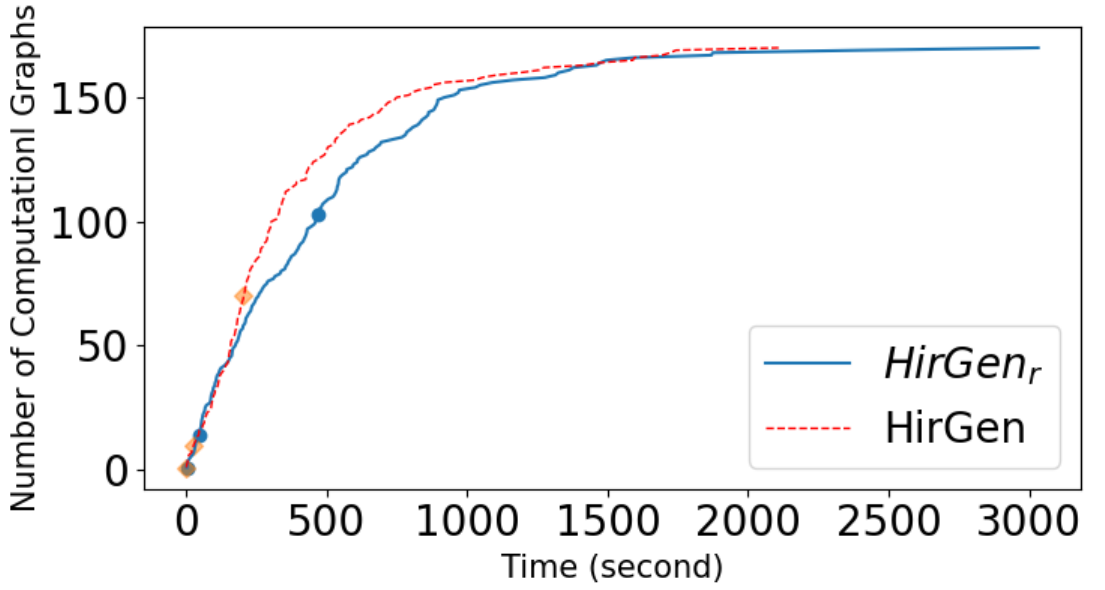}
    \caption{\techname{} \textit{vs.} \techname{}$_{r}$ under Disruptive Generation}
    \label{figure:vsdisruptive}
\end{figure}

The second experiment is similar to the first one. In this experiment, we compare the bug detection ability of \techname{} and \techname{}$_{r}$ under the disruptive generation mode. Since disruptive generation promises that each insertion contains a violation of constraints and must trigger failure, there is no need to generate a multiple-operator graph. So we let them generate one-node graphs. Figure \ref{figure:vsdisruptive} presents the experiment results. \techname{} and \techname{}$_{r}$ both generated 170 bug-triggered computational graphs, each of which contains unique tuples of (\textit{operator}, \textit{tensor shape}, \textit{data type}). In this figure, \techname{} shows a more exploratory nature in the diversity of graphs and thus detects bugs faster. Further, two techniques both found 3 bugs using these 170 graphs. And the timestamps of bug detection are also marked in this figure, showing that \techname{} found bugs faster than \techname{}$_{r}$.


\section{Discussion}
\label{sec:6}

\subsection{Limitation and Benefit of $Oracle_3$}
\label{sec:oracle3_limits}
In our experiment, $Oracle_3$ is less effective than others. It only detected three confirmed but not fixed bugs, implying that 1) it's not as efficient as other oracles in bug detection, and 2) fix of these bugs is of low priority from the perspective of developers. 
The reason for this phenomenon is that the floating-point precision settings differ across platforms~\cite{MTXiao}.
And it is hard for developers to tell whether the differences are caused by bugs or inconsistent precision settings.
Therefore, they are reluctant in checking the code or refining the exception-handling module to warn users of the large differences in calculation results among platforms when they use some special operators. 
Despite this limitation, $Oracle_3$ could still help find confusing scenarios and provide experience for users who do not have enough knowledge of special error-triggering operators in low-level platforms. Take $Bug_{15}$~\cite{TVM-bug} as an example. 
If TVM compiles the computational graph construction including {\tt right\_shift}, then LLVM may skew the results. Since TVM offers no warning, this phenomenon is confusing. We posted such findings online and obtained such a response from a developer: \textit{There is not a full specification for right\_shift intrin in TVM}. Therefore, we counted it as a TVM bug of incomplete documentation and exception handling module.

\subsection{Threats to Validity}
The threat to \textit{internal} validity mainly lies in the implementation of \techname{}. To reduce this threat, two authors of this paper have carefully checked and tested the functionality of all components of \techname{}.

The threat to \textit{external} validity mainly lies in the DL compiler we chose in our study. Until now, TVM is one of the most popular and active open-source DL compilers, with 9K stars on GitHub. 
Though \techname{} now mainly supports converting its generated computational graph into the high-level IR of TVM with Relay. 
The technical approach is also useful for testing other DL compilers with the help of ONNX\cite{onnx}. 
ONNX is an open format to represent diverse DL models defined by various DL frameworks and is currently supported by popular DL compilers. Similar to Relay, we can use ONNX's APIs to easily convert a computational graph into a high-level IR of ONNX. This IR is transformable to high-level IRs of existing DL compilers. Adding more support for ONNX to test more DL compilers is also our future work.

The threat to \textit{construct} validity mainly lies in randomness and settings. In computational graph generation, though with coverage guidance, the selection of operator and connection also involves randomness. To alleviate the negative impact of randomness, we 1) repeated all randomness-involved experiments 10 times and utilize average, variance, and Mann-Whitney U Test to promise the results are statistically significant.
The threshold setting for comparing different prediction results in $Oracle_2$ and $Oracle_3$ (mainly $Oracle_3$) is still an open problem in DL compiler testing. One existing work~\cite{MTXiao} has shown that different floating-point precision settings in different platforms may lead to false positives in bug detection. Therefore, in comparing prediction results, threshold setting is vital in reducing false positives. Since no systematic study on how to set threshold exists, our settings are mainly based on experience and expertise in testing TVM. Since our test oracles did not frequently detect prediction differences, we set the threshold to a tiny floating number, $10^{-3}$, to not miss any minor difference, and thus not miss any new bug. In the experiment, \techname{} did not find false positives due to the floating-point roundoff. 




\section{Related Work}
\subsection{DL Compiler Testing}
\label{sec:NNSmith}
With the development of DL compiler, the importance of DL compiler testing has been noticed by more and more researchers. 
According to the testing focus,
existing testing techniques can be divided into two categories. 
The first category aims at testing the whole workflow of DL compilers. 
Focusing on testing a single stage is another category.
MT-DLComp and NNSmith are the former.
MT-DLComp~\cite{MTXiao} can perform semantics-preserving mutation on seed DL models to generate new models with theoretically the same prediction capability. During testing, any prediction difference between mutated models and the seed model or any compilation failure will be captured.
NNSmith~\cite{NNSmith} can generate computational graphs and their inputs/weights from scratch to test DL compilers. 
Although NNSmith and \techname{} both generate computational graphs, 
\techname{} are fundamentally different from NNSmith at least in terms of their testing purposes, their usage of the generated computational graphs in DL compiler testing, and the types of bugs detected.
As for testing purposes, \techname{} focuses on the most error-prone stage~\cite{qingchao}, \ie,
the high-level optimization stage, while NNSmith has no testing preference for any compilation stage and focuses on validating the prediction correctness of the compiled models. Though \techname{} can also cover low-level and codegen components, most of its technical details, including mutation strategies, use of high-level optimizations, and construction of $Oracle_2$, are all designed for the only stage.
With different testing purposes, they utilize computational graphs differently. 
\techname{} further generates multiple semantics-equivalent high-level IRs from the computational graphs and uses high-level optimizations to optimize them. 
NNSmith does not perform these steps.
The utilization of high-level IRs and optimizations helps \techname{} find much more high-level optimization bugs than NNSmith.
NNSmith further finds a set of inputs and weights for the computational graphs such that the compiled DL models produce numerically valid outputs given such inputs and weights. Then NNSmith can validate prediction results and catch any prediction errors.
Though \techname{} could also cause several prediction errors, it's not as efficient as NNSmith.
Because of these differences, 
\techname{} and NNSmith has nearly orthogonal bug detection ability, as shown in section \ref{SOTA}.

Different from MT-DLComp and NNSmith, several other techniques~\cite{liu2022coverage,TVMFuzz,qingchao} focus on the testing of a single stage but not the whole workflow. Besides, they perform white-box testing to utilize knowledge gained from the codebase to achieve more efficient and effective testing results. For instance, TZER\cite{liu2022coverage} collects low-level IR passes and mutates them to detect bugs in low-level optimizations,
while TVMfuzz~\cite{qingchao} focuses on high-level optimization by generating high-level API sequences.

\vspace{-2mm}
\subsection{Metamorphic Testing for Compiler}
Metamorphic testing (MT)~\cite{metamorphic} is a popular approach to address the test oracle problem. 
Researchers proposed different metamorphic relations (MR) to construct test oracles for different systems under test based on the characteristics of the systems. 

 In compiler testing, the follow-up test inputs in MR are mostly programs equivalent to their seed programs~\cite{compilertestingsurvey}. Various semantics-preserving mutations have been proposed to generate equivalent programs for compiler testing~\cite{EMI, Athena, Hermes, MTXiao, shader}. 
MT-DLComp~\cite{MTXiao} conducts metamorphic testing on DL compilers via two semantics-preserving mutations on computational graphs. Specifically, it inserts always-yield-zero nodes into computational graphs to generate new graphs without skewing the calculation. 
GLFuzz~\cite{shader} tests OpenGL by designing six semantics-preserving mutators.
Unlike MT-DLComp, $Oracle_2$ of \techname{} is for high-level IRs instead of computational graphs, and semantics-preserving mutations in \techname{} are also for high-level IRs. 
Moreover, the mutations of \techname{} focus on modifying the function call chain, which is orthogonal to all six mutations in GLFuzz.
EMI~\cite{EMI}, which stands for equivalence modulo inputs, is a methodology for constructing MRs~\cite{compilertestingsurvey}. 
The key insight is that given a seed program, a set of inputs can induce a collection of programs giving the same outputs as the seed one on these inputs. 
EMI has inspired several compiler testing techniques~\cite{EMI, many-core, Athena, Hermes}. The semantics-preserving mutation in EMI-based techniques requires profiling program executions before mutation. In addition, EMI-based mutants are constructed to be semantics-equivalent only under specific inputs. In contrast, \techname{} requires no profiling, and the mutants generated are semantically equivalent to their seed high-level IRs under all inputs.

\section{Conclusion}
High-level optimization is the most bug-prone stage in the workflow of DL compilers. 
However, there is no systematic study on testing this stage. To fill this gap, we offer \techname{}, a generation-based fuzzer with an effective computational graph generation approach and three test oracles. Different from existing works, \techname{} can explore more complicated and valid high-level IRs and thus detect deeper bugs. Besides, three test oracles in \techname{} also improve its capability of detecting bugs of various root causes. 
Our effort improved the robustness and functional correctness of high-level optimization and was recognized by the TVM community.

\section*{Acknowledgement}
This work is supported by the National Natural Science Foundation of China (Grant No. 61932021), Hong Kong Research Grant Council/General Research Fund (Grant No.
16205722), Hong Kong Research
Grant Council/Research Impact Fund (Grant No. R5034-18), and National Natural Science Foundation of China under Grant Nos.62002256 and 62232001.

\balance
\bibliographystyle{ACM-Reference-Format}
\bibliography{reference}

\end{document}